\pdfoutput=1

\documentclass[11pt]{article}

\usepackage[preprint]{acl}
\usepackage{times}
\usepackage{latexsym}

\usepackage[T1]{fontenc}

\usepackage[utf8]{inputenc}

\usepackage{microtype}

\usepackage{inconsolata}

\usepackage{amsmath}
\usepackage{graphicx}
\usepackage{alltt}
\usepackage{geometry}
\usepackage{multicol}
\usepackage{listings}     
\usepackage[most]{tcolorbox}
\usepackage{lstautogobble}  
\usepackage{color}          
\usepackage{xspace}
\usepackage{ulem}
\usepackage{booktabs}  
\usepackage{xcolor}    
\usepackage{graphicx}  
\usepackage{multirow}  
\usepackage{colortbl}  
\usepackage{enumitem}

\usepackage{float} 
\definecolor{darkgreen}{HTML}{006400}
\definecolor{darkred}{HTML}{8B0000}
 
\newcommand{\tool}{\textsc{Llm4Effi}\xspace}

\title{\tool: Leveraging Large Language Models to Enhance \\ Code Efficiency and Correctness}

\author{
Tong Ye\textsuperscript{1}, 
Weigang Huang\textsuperscript{1}, 
Xuhong Zhang\textsuperscript{1}, 
Tengfei Ma\textsuperscript{2},  \\
\bf{Peiyu Liu}\textsuperscript{1},
\bf{Jianwei Yin}\textsuperscript{1},
\bf{Wenhai Wang}\textsuperscript{1}\\
        \textsuperscript{1}Zhejiang University; \\
        \textsuperscript{2}Stony Brook University\\
        \texttt{\{{tongye,huangweigang,zhangxuhong,liupeiyu,zdzzlab}\}@zju.edu.cn} \\
        \texttt{zjuyjw@cs.zju.edu.cn}, 
        \texttt{tengfei.ma@stonybrook.edu}
        }

\begin{document}
\maketitle
\begin{abstract}
Large Language Models (LLMs), particularly Code LLMs, have demonstrated impressive performance in code generation. Current research primarily focuses on the correctness of generated code, while efficiency remains less explored. Recent works have focused on modifying the initial version of the code to improve its efficiency. However, such refinements are limited by the algorithmic design and overall logic of the initial code, resulting in only incremental improvements. In contrast, when human developers write high-quality code, they typically begin by designing several potential solutions at the logical level, evaluating various algorithms and their complexities, and then proceeding to implement and optimize the solution. In this study, we introduce \tool: \uline{L}arge \uline{L}anguage \uline{M}odel for Code \uline{Effi}ciency, a novel framework that enables LLMs to generate code that balances both efficiency and correctness. Specifically, \tool divides the efficiency optimization process into two domains: algorithmic exploration in the logic domain and implementation optimization in the code domain. The correctness of the code is then guaranteed through a synthetic test case refinement process. This approach, which prioritizes efficiency before ensuring correctness, offers a new paradigm for efficient code generation. Experiments demonstrate that \tool consistently improves both efficiency and correctness, achieving new state-of-the-art performance in code efficiency benchmarks across various LLM backbones.
\end{abstract}

\section{Introduction.}
Large Language Models (LLMs), particularly those specialized in code, are revolutionizing the field of software engineering at an unprecedented pace. A significant area of advancement lies in automated code generation \cite{evalplus}, where LLMs such as GPT-4o \cite{openai2024gpt4ocard}, Gemini \cite{team2023gemini}, the DeepSeek Series \cite{deepseekv2}, and the Qwen Series \cite{qwen2, qwen2.5} demonstrating remarkable capabilities. These models have attracted considerable attention from both academia and industry, consistently breaking new ground on code completion and generation benchmarks, including HumanEval \cite{chen2021evaluating}, MBPP \cite{austin2021program}, and LiveCodeBench \cite{jain2024livecodebench}.

\begin{figure}
    \centering
    \includegraphics[width=1.0\linewidth]{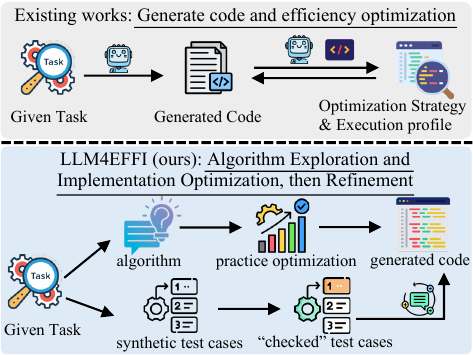}
    \caption{Comparison of \tool with existing methods. Existing methods generate code first, then optimize it using strategy and execution profiles. In contrast, \tool starts with the task, focusing on efficiency through algorithm exploration and implementation, followed by correctness refinement.}
    \label{fig:comparison}
\end{figure}

While these LLMs achieve impressive accuracy in automatic code generation, practical software engineering applications require more than just correct code—they also demand efficiency \cite{shi2024efficientgreenlargelanguage,niu2024evaluatingefficiencysourcecode}. In real-world scenarios, even correct code often requires manual optimization by engineers before it can be used, which undermines the goal of "out-of-the-box" automated code generation. Therefore, generating code that is both correct and efficient is essential, yet automating this process has not been widely explored.

\begin{figure*}
    \centering
    \includegraphics[width=1.0\textwidth]{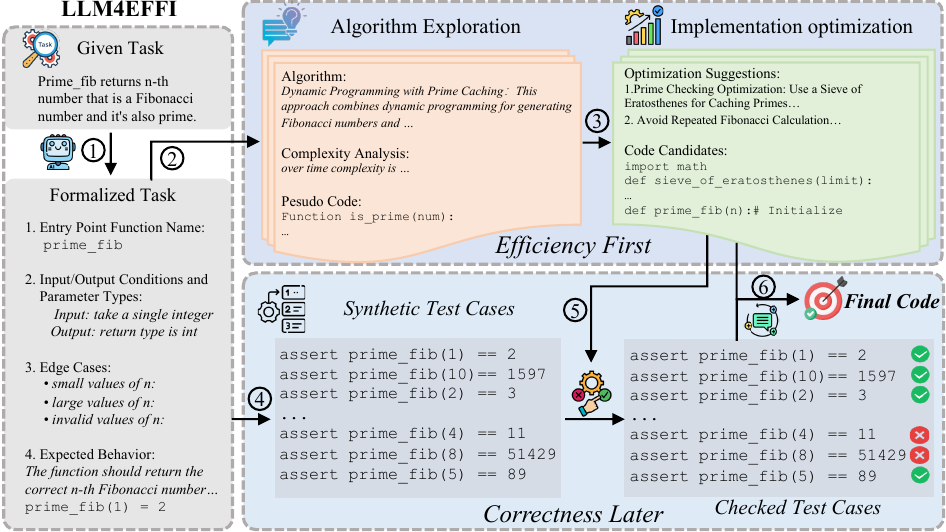}
    \caption{The workflow of \tool. Given a programming task, \tool formalizes it into a code-oriented description, generates optimal algorithms and pseudocode in logic domain, and then produces implementation suggestions in code domain. \tool synthesizes test cases and uses a verification-based adaptive framework to evaluate candidate solutions. The final code is selected based on the highest pass rate of the "checked" test cases.}
    \label{fig:workflow}
\end{figure*}

Recent preliminary works \cite{EffiLearner,waghjale-etal-2024-ecco} have explored feedback-based approaches to optimize generated code and enhance its efficiency. As illustrated in Figure \ref{fig:comparison}, these methods typically involve profiling code execution time and incorporating reflective feedback into the optimization process. However, the "generate-then-optimize" paradigm is constrained by the algorithmic design and overall structure of the initial code, leading to only incremental improvements. We provide detailed examples in Figure \ref{fig:case1} and \ref{fig:case2} in Appendix B. In contrast, when human developers write high-quality code, whether in practical software development or algorithmic teaching scenarios, they typically start by designing multiple potential solutions at a logical level. For example, when tackling a sorting problem, a developer might consider Quicksort for its average-case efficiency of $\mathcal{O}(N*\log N)$, while also factoring in its worst-case time complexity of $\mathcal{O}(N^2)$. By carefully analyzing the problem’s constraints and evaluating various algorithms along with their complexities, they then proceed to implement the solution, applying various coding techniques to optimize it. Finally, they debug and refine the code to achieve a high-quality implementation.

Inspired by this thought, we propose \tool, as shown in Figure \ref{fig:workflow}, a novel paradigm that enables LLMs to generate both efficient and correct code. Specifically, for a given programming task described in natural language, \tool first "formalizes" it into a code-oriented problem description. In other words, it converts the broad natural language statement into a clear, concrete, and well-defined coding problem, ensuring that the LLM can accurately interpret it. Next, \tool prompts the LLM for logic-level reasoning and exploration, considering various algorithmic approaches, providing corresponding complexity analyses, and generating relevant pseudocode. Based on these different algorithm designs and their associated pseudocode, \tool suggests code implementation strategies, followed by code generation and optimization at the implementation level, as high-quality code also requires careful consideration during the practical implementation stage. To ensure the functional correctness of the generated code while targeting efficiency, \tool introduces a bidirectional verification-based adaptive testing framework to check synthetic test cases. Finally, the code solutions are executed on the "checked" test cases and iterated upon for correctness. The solution with highest pass rate across the "checked" test cases is selected as the final generated code.

The \tool has two distinctive uniqueness:
\noindent\textbf{\underline{Uniqueness 1:}} Separation of Efficiency Optimization into Logic and Code Domains. \tool divides efficiency optimization into two distinct domains: the "logic domain" and the "code domain". In the logic domain, efficiency optimization focuses on exploring the optimal algorithmic approaches, while in the code domain, optimization deals with the practical implementation details. This separation effectively breaks down the challenge of optimizing code efficiency into manageable steps, making the overall efficiency optimization process more systematic and targeted.

\noindent\textbf{\underline{Uniqueness 2:}} The Order of Correctness and Efficiency. The order in which correctness and efficiency are optimized plays a critical role. By prioritizing efficiency first, a wider range of algorithmic solutions can be explored, leading to the discovery of multiple efficient approaches. Correctness is then incrementally ensured across these solutions. This approach avoids prematurely constraining efficiency optimization by focusing too early on correctness. Prioritizing efficiency first allows for greater room for improvement and significantly enhances the potential for efficiency gains.

We validate \tool on three recently proposed code efficiency benchmarks: EvalPerf \cite{liu2024evaluatinglanguagemodelsefficient}, ENAMEL \cite{qiu2024efficientllmgeneratedcoderigorous}, and Mercury \cite{du2024mercury}. Experimental results show that \tool consistently enhances both code correctness and efficiency across various LLM backbones, achieving state-of-the-art performance in efficiency metrics. Specifically, using the DeepSeek-V3 backbone, \tool improves eff@1 by 9.27\% on ENAMEL and boosts DPS\_norm by 6.63\% on Mercury.

Overall, we summarize our contributions as follows, with corresponding code available at link\footnote{\url{https://anonymous.4open.science/r/LLM4EFFI-04B2}}:
\begin{itemize}[itemsep=2pt,topsep=0pt,parsep=0pt]
\item We propose \tool, the first framework that simultaneously optimizes both code efficiency and correctness.
\item We introduce two key features: \textit{Separation of Efficiency Optimization into Logic and Code Domains} and \textit{Order of Correctness and Efficiency}. We hope these unique features will contribute to the advancement of the code efficiency community.
\item Extensive experiments and analysis on three benchmarks across different LLM backbones demonstrate the effectiveness and robustness of \tool in efficient code generation.
\end{itemize}

\section{Related Works}
\subsection{LLMs for Code Domain.}
Large language models have been widely applied to coding tasks and have shown strong performance across various coding scenarios and evaluations. Most existing research focuses on code generation, with numerous techniques developed to enhance its quality. Some methods aim to improve the quality of synthetic code data \cite{wei2024magicoder, luo2024wizardcoder, lei2024autocoder}, enhance self-consistency \cite{le2024codechain, huang-etal-2024-enhancing}, or leverage feedback from human or LLM annotations \cite{chen2024improvingcodegenerationtraining, wu2023finegrained, tang-etal-2023-explain}. Other approaches utilize multi-agent collaboration frameworks to enhance code generation \cite{zhong2024debug, shinn2023reflexion, islam-etal-2024-mapcoder, madaan2023selfrefine, li2024codetreeagentguidedtreesearch}. However, these methods primarily focus on the correctness of the generated code, with relatively little emphasis on the efficiency of the generated code.

\subsection{Code Efficiency.}
Until recently, the academic community has only begun to pay significant attention to the efficiency of generated code. Recently, several efficiency-focused benchmarks \cite{EffiBench, du2024mercury, liu2024evaluatinglanguagemodelsefficient, qiu2024efficientllmgeneratedcoderigorous} have emerged, aiming to provide a more comprehensive evaluation of LLMs’ ability to generate efficient code. However, empirical studies and evaluations of these benchmarks show that current LLMs still face significant challenges in generating efficient code. To improve code efficiency, recent research such as ECCO \cite{waghjale-etal-2024-ecco} adopts self-refinement, prompting LLMs to consider possible optimization strategies and refine their outputs. Effi-Learner \cite{EffiLearner} proposes a self-optimization framework that uses execution overhead profiles, feeding them back into the LLM to revise the code and reduce time overhead. However, these methods focus on enhancing the efficiency of code after it has been generated, rather than starting with the goal of generating both efficient and correct code from the beginning.

\section{Methodology}
\paragraph{Problem Formulation.}
In the code efficiency task, each sample is represented as a pair $(Q, T_h)$, where $Q$ denotes the task description, and $T_h$ corresponds to the hidden test cases. Our goal is to generate the corresponding code solution $S$ that passes the hidden test cases and achieves the highest efficiency (i.e., the shortest execution time). Notably, to better simulate real-world scenarios, we assume there are no public test cases. $T_h$ is only used during the evaluation stage and is not visible during efficiency and correctness optimization stages.

\subsection{Overview.}
We present the framework of \tool in Figure \ref{fig:workflow}. For a given programming task described in natural language, \tool first "formalizes" it into a code-oriented problem description (\textbf{Section \ref{method-1}}). Next, \tool queries the LLM for logic-domain reasoning and exploration, generating multiple optimal algorithmic solutions along with their corresponding pseudocode (\textbf{Section \ref{method-2}}). Based on these algorithm designs and their associated pseudocode, \tool analyzes and generates code implementation suggestions, followed by the generation and optimization of the corresponding code at the implementation level (\textbf{Section \ref{method-3}}). To further refine the solutions for correctness, \tool synthesizes a large number of test cases and utilizes a bidirectional verification-based adaptive testing framework to "check" these synthetic test cases. The "checked" test cases are then used to evaluate the candidate code solutions (\textbf{Section \ref{method-4}}). The solution with highest pass rate across the "checked" test cases is selected as the final generated code.

\subsection{Task Formalization.}
\label{method-1}
In the initial task formalization stage, \tool ensures that the task description is clear and unambiguous, which is crucial for the success of subsequent stages. As highlighted by \citet{han-etal-2024-archcode}, errors in LLM-generated code often arise from an insufficient or unclear understanding of the task. Therefore, \tool prompts the LLM to comprehend the task from four key dimensions: \uline{entry point function name}, \uline{input/output conditions and parameter types}, \uline{edge cases}, and \uline{expected behavior}. Based on these dimensions, the LLM is further encouraged to engage in self-reflection to confirm whether it has fully grasped all aspects of the task, thus laying a solid foundation for the subsequent stages. Formally, $Q \to Q_{formal} \overset{\text{check}} {\longleftrightarrow} Q$.

\subsection{Algorithmic Exploration in Logic Domain.}
\label{method-2}
For the formalized task defined in the first stage, \tool prompts the LLM to engage in algorithmic reasoning at the logical level, rather than immediately generating code. This approach mirrors that of human programmers, who first perform abstract and high-level reasoning before implementation. The LLM is prompted to explore multiple potential optimal algorithms, analyze their corresponding complexities, and represent the entire logical process with pseudocode. Formally, $Q_{formal} \to \{Algo, Cplx, Pseudo\}$, where $Algo$ refers to the algorithm plan, $Cplx$ refers to the complexity analysis, and $Pseudo$ refers to the corresponding pseudocode.

\subsection{Implementation Optimization in Code Domain.}
\label{method-3}
Excellent code not only requires careful algorithm design but also necessitates optimization at the implementation level. Even when the same algorithm is used, different implementation approaches can lead to significant variations in code efficiency \cite{shypula2024learning, Coignion_2024}. When implementing code based on the algorithm plan and corresponding pseudocode, \tool prompts the LLM to provide practical suggestions derived from $Algo$ and $Pseudo$, such as replacing a manual binary exponentiation implementation with Python’s built-in pow function, among other optimizations. We provide three detailed examples in the appendix to illustrate this process. Subsequently, \tool generates the corresponding code based on the $Algo$, $Pseudo$, and implementation suggestions, while also checking for further optimization opportunities. Formally, 
$\{Algo, Pseudo\}\to\{Suggs\}$, and $\{Algo, Pseudo, Suggs\}\to\{Code \ Candidates\}$.

\begin{table*}[ht]
    \centering
    \small
    \renewcommand{\arraystretch}{1.1}
    \resizebox{\textwidth}{!}{%
    \begin{tabular}{ll cc cc cc}
\toprule
\multirow{2}{*}{\textbf{LLMs}} & \multirow{2}{*}{\textbf{Methods}} & \multicolumn{2}{c}{\textbf{EvalPerf}} & \multicolumn{2}{c}{\textbf{Mercury}} & \multicolumn{2}{c}{\textbf{ENAMEL}} \\
\cmidrule(lr){3-4} \cmidrule(lr){5-6} \cmidrule(lr){7-8}
 &  & \textbf{DPS\_norm} & \textbf{Pass@1} & \textbf{Beyond@1} & \textbf{Pass@1} & \textbf{eff@1} & \textbf{Pass@1} \\
\midrule
\midrule
\multirow{4}{*}{\shortstack{Qwen2.5-Coder\\-32B-Instruct}}
& Instruct    & 80.92 & 85.59 & 76.97 & \textbf{94.14} & 50.44 & 85.21 \\
& ECCO       & 82.16 & 63.56 & 73.29 & 89.06 & 41.89 & 71.83 \\
& Effi-Learner & 82.45 & 77.11 & 77.13 & 91.41 & 50.12 & 81.69 \\
& \textbf{\tool (ours)}     & \textbf{86.20 } \textcolor{darkgreen}{ +5.28} & \textbf{87.30 } \textcolor{darkgreen}{ +1.71} & \textbf{78.96 } \textcolor{darkgreen}{ +1.99} & 93.75 \textcolor{darkred}{-0.39} & \textbf{51.26 } \textcolor{darkgreen}{ +0.82} & \textbf{86.62 } \textcolor{darkgreen}{ +1.41} \\
\midrule

\multirow{4}{*}{\shortstack{Qwen2.5-72B\\-Instruct}}
& Instruct   & 79.29 & 88.14 & 72.50 & 86.72 & 49.78 & 83.80 \\
& ECCO      & 80.06 & 64.41 & 74.10 & 89.84 & 41.90 & 72.53 \\
& Effi-Learner & 79.90 & 81.36 & 77.10 & \textbf{91.02} & 47.42 & 76.76 \\
& \textbf{\tool (ours) }     & \textbf{84.00 } \textcolor{darkgreen}{ +4.71} & \textbf{88.98 } \textcolor{darkgreen}{ +0.84} & \textbf{77.45 } \textcolor{darkgreen}{ +4.95} & 90.63 \textcolor{darkgreen}{ +3.91} & \textbf{51.49 } \textcolor{darkgreen}{ +1.71} & \textbf{87.32 } \textcolor{darkgreen}{ +3.52} \\
\midrule

\multirow{4}{*}{GPT-4o-mini} 
& Instruct   & 80.04 & 85.59 & 69.59 & 82.81 & 48.26 & 80.28 \\
& ECCO      & 75.18 & 44.07 & 72.29 & 86.33 & 30.75 & 57.75 \\
& Effi-Learner & 79.80 & 81.36 & 73.45 & 88.67 & 45.69 & 77.46 \\
& \textbf{\tool (ours)}      & \textbf{83.78 } \textcolor{darkgreen}{ +3.74} & \textbf{88.14 } \textcolor{darkgreen}{ +2.55} & \textbf{74.94 } \textcolor{darkgreen}{ +5.35} & \textbf{89.45 } \textcolor{darkgreen}{ +6.64} & \textbf{49.89 } \textcolor{darkgreen}{ +1.63} & \textbf{80.99 } \textcolor{darkgreen}{ +0.71} \\
\midrule

\multirow{4}{*}{GPT-4o}
& Instruct   & 79.59 & 86.70 & 73.14 & 87.50 & 47.63 & 80.99 \\
& ECCO      & 80.65 & 61.02 & 77.70 & 92.18 & 38.63 & 64.79 \\
& Effi-Learner & 79.39 & 79.67 & \textbf{79.24} & 93.36 & 48.52 & 81.69 \\
& \textbf{\tool (ours)}      & \textbf{86.39 } \textcolor{darkgreen}{ +6.80} & \textbf{88.98 } \textcolor{darkgreen}{ +2.28} & 77.81 \textcolor{darkgreen}{ +4.67} & \textbf{93.75 } \textcolor{darkgreen}{ +6.25} & \textbf{55.26 } \textcolor{darkgreen}{ +7.63} & \textbf{83.80 } \textcolor{darkgreen}{ +2.81} \\
\midrule

\multirow{4}{*}{DeepSeek-V3} 
& Instruct   & 80.45 & 89.84 & 79.90 & 94.53 & 51.14 & 86.62 \\
& ECCO      & 81.08 & 61.84 & 63.26 & 74.61 & 45.84 & 75.35 \\
& Effi-Learner & 79.00 & 88.14 & 78.83 & 92.58 & 52.22 & 83.80 \\
& \textbf{\tool (ours)}      & \textbf{87.08 } \textcolor{darkgreen}{ +6.63} & \textbf{90.67 } \textcolor{darkgreen}{ +0.83} & \textbf{82.76 } \textcolor{darkgreen}{ +2.86} & \textbf{96.09 } \textcolor{darkgreen}{ +1.56} & \textbf{60.41 } \textcolor{darkgreen}{ +9.27} & \textbf{89.44 } \textcolor{darkgreen}{ +2.82} \\
\bottomrule
    \end{tabular}%
    }
    \caption{\textbf{Main Result.} The results of \tool, Instruct, ECCO, and Effi-Learner methods on the EvalPerf, Mercury, and ENAMEL benchmarks are presented using the Qwen2.5-Coder-32B-Instruct, Qwen2.5-72B-Instruct, GPT-4o-mini, GPT-4o, and DeepSeek-V3 LLM backbones. Correctness is evaluated using Pass@1, while efficiency is measured using the respective efficiency metrics for each of the three benchmarks.}
    \label{tab:complete_results}
\end{table*}

\subsection{Code Correctness.}
\label{method-4}
To ensure the functional correctness of generated code while targeting efficiency, \tool introduces a bidirectional verification-based adaptive testing framework. The process works as follows: First, \tool automatically synthesizes a large number of test cases based on the formalized task description $Q_{formal}$. These test cases are designed to cover a wide range of edge cases, thoroughly testing the robustness and reliability of the generated code. However, since the synthesized test cases may not be entirely correct, \tool performs bidirectional verification to validate them.

\noindent \textbf{Forward Verification:} If all candidate code implementations pass a specific test case, the test case is marked as trusted. Otherwise, \textbf{Reverse Review:} For test cases that cause failures in any candidate code, \tool performs a $Q_{formal}$-based review. It checks whether the test case aligns with the intent of the formal task description and conducts a semantic consistency check, which is similar to Test-Driven Development \cite{Erdogmus2010TestDrivenD} in software engineering. If the test case passes the reverse review, it is retained; otherwise, it is discarded. Finally, the retained test cases are marked as "checked". These "checked" test cases are then used to evaluate the generated code candidates, with any failures triggering further refinements. The code candidate that passes the most "checked" test cases is ultimately selected as the final solution. Formally,
$Q_{formal} \to \{Synth. \ test \ cases\} $$\overset{\text{check}} \to \{Checked\ test\ cases\} \overset{\text{select}} \to \{final \ solution\}$.

\section{Experiments}
\begin{table*}[ht]
    \centering
    \small
    \renewcommand{\arraystretch}{1.1}
    \resizebox{\textwidth}{!}{%
    \begin{tabular}{ll cc cc cc}
        \toprule
        \multirow{2}{*}{\textbf{Models}} & \multirow{2}{*}{\textbf{Methods}} & \multicolumn{2}{c}{\textbf{EvalPerf}} & \multicolumn{2}{c}{\textbf{Mercury}} & \multicolumn{2}{c}{\textbf{ENAMEL}} \\
        \cmidrule(lr){3-4} \cmidrule(lr){5-6} \cmidrule(lr){7-8}
         &  & \textbf{DPS\_norm} & \textbf{Pass@1} & \textbf{Beyond@1} & \textbf{Pass@1} & \textbf{eff@1} & \textbf{Pass@1} \\
        \midrule
        \midrule

        \multirow{4}{*}{\shortstack{Qwen2.5-Coder\\-32B-Instruct}}
        & \tool          & \textbf{86.20} & \textbf{87.30} & \textbf{78.96} & \textbf{93.75} & \textbf{51.26} & \textbf{86.62} \\
         & Variant-1      & 77.21 \textcolor{darkred}{-8.99} & 80.51 \textcolor{darkred}{-6.79} & 77.89 \textcolor{darkred}{-1.07} & 93.34 \textcolor{darkred}{-0.41} & 48.57 \textcolor{darkred}{-2.69} & 81.69 \textcolor{darkred}{-4.93} \\
         & Variant-2      & 75.75 \textcolor{darkred}{-10.45} & 81.36 \textcolor{darkred}{-5.94} & 75.86 \textcolor{darkred}{-3.10} & 92.19 \textcolor{darkred}{-1.56} & 45.68 \textcolor{darkred}{-5.58} & 83.10 \textcolor{darkred}{-3.52} \\
         & Variant-3      & 81.19 \textcolor{darkred}{-5.01} & 72.03 \textcolor{darkred}{-15.27} & 72.56 \textcolor{darkred}{-6.40} & 85.16 \textcolor{darkred}{-8.59} & 47.48 \textcolor{darkred}{-3.78} & 77.46 \textcolor{darkred}{-9.16} \\

        \midrule
  \multirow{4}{*}{DeepSeek-V3}
           & \tool          & \textbf{87.08} & \textbf{90.67} & \textbf{82.76} & \textbf{96.09} & \textbf{60.41} & \textbf{89.44} \\
         & Variant-1      & 79.72 \textcolor{darkred}{-7.36} & 84.75 \textcolor{darkred}{-5.92} & 81.58 \textcolor{darkred}{-1.18} & 94.53 \textcolor{darkred}{-1.56} & 53.23 \textcolor{darkred}{-7.18} & 88.03 \textcolor{darkred}{-1.41} \\
         & Variant-2      & 77.07 \textcolor{darkred}{-10.01} & 83.05 \textcolor{darkred}{-7.62} & 80.10 \textcolor{darkred}{-2.66} & 94.53 \textcolor{darkred}{-1.56} & 53.62 \textcolor{darkred}{-6.79} & 88.73 \textcolor{darkred}{-0.71} \\
         & Variant-3      & 82.62 \textcolor{darkred}{-4.46} & 82.01 \textcolor{darkred}{-8.66} & 79.75 \textcolor{darkred}{-3.01} & 92.58 \textcolor{darkred}{-3.51} & 54.58 \textcolor{darkred}{-5.83} & 81.69 \textcolor{darkred}{-7.75} \\

        \bottomrule
    \end{tabular}
    }
    \caption{\textbf{Ablation Study Results.} The results of \tool, Variant-1, Variant-2, and Variant-3 are presented using Qwen2.5-Coder-32B-Instruct and DeepSeek-V3 as LLM backbones on the EvalPerf, Mercury, and ENAMEL benchmarks. We have highlighted the performance changes of Variants compared to \tool with colors.}
    \label{tab:ablation_results}
\end{table*}
We evaluate \tool on three code efficiency evaluation benchmarks: EvalPerf \cite{liu2024evaluatinglanguagemodelsefficient}, Mercury \cite{du2024mercury} and ENAMEL \cite{qiu2024efficientllmgeneratedcoderigorous}. \textbf{\uline{EvalPerf}} focuses on performance-challenging tasks and uses Differential Performance Evaluation to assess efficiency across different LLMs and solutions. Its efficiency metric, DPS\_norm, is calculated by determining the cumulative ratio of the reference solution that is immediately slower than the new solution, normalized by the total number of solutions. This ensures a fair comparison of code efficiency based on reference solutions with varying performance levels. \textbf{\uline{Mercury}} introduces the Beyond metric to evaluate both functional correctness and code efficiency. The Beyond metric is calculated by normalizing the runtime percentiles of LLM solution samples over the runtime distribution for each task, ensuring consistent runtime comparisons across different environments and hardware configurations. \textbf{\uline{ENAMEL}} evaluates code efficiency using the eff@1 metric. This efficiency score is determined by measuring the worst execution time of the code sample across test cases of varying difficulty levels. The score is then adjusted using a weighted average across these levels to account for hardware fluctuations. The eff@1 metric ranges from 0 to 1, with higher values indicating greater code efficiency. A value exceeding 1 signifies that the generated code is more efficient than the expert-level solution.


\subsection{Compared Methods.}
We evaluate the direct instruction of generating correct and efficient code as the \textbf{Instruct} baseline. We compare \tool with two recent proposed methods \textbf{ECCO} \cite{waghjale-etal-2024-ecco} and \textbf{Effi-Learner} \cite{EffiLearner} for code efficiency. 
\begin{itemize}[leftmargin=*,itemsep=2pt,topsep=0pt,parsep=0pt]
\item \textbf{ECCO:} A self-refine with NL feedback approach that prompts the LLM to generate code, then asks if improvements in correctness or efficiency can be made, and finally refines the solution based on optimization suggestions.
\item \textbf{Effi-Learner:} First generates code using instruction prompts same as \textbf{Instruct} baseline, then executes the code with test cases to collect performance profiles, including runtime and memory usage. These profiles are fed back into the LLM along with the code, prompting the LLM to refine the code for efficiency based on the profile. It is worth noting that Effi-Learner relies on test case oracles, and in this study, we use the visible test cases from the task. In contrast, \tool does not rely on any test case oracles; all test cases are synthetically generated by \tool itself.
\end{itemize}

\subsection{Experiment Setup.}
To comprehensively evaluate \tool, we selected five different LLM backbones: two proprietary models, GPT-4o \cite{openai2024gpt4ocard} and GPT-4o-mini, and three open-source models, including DeepSeek-V3  \cite{deepseekai2024deepseekv3technicalreport}, Qwen2.5-72B-Instruct \cite{qwen2.5}, and Qwen2.5-Coder-32B-Instruct \cite{hui2024qwen2}. During the \tool process, we set the number of algorithm plans to 5 and the number of synthetic test cases to 20, followed by one iteration to refine the code for correctness. All prompts used in \tool are detailed in Appendix \ref{sec:ap-method}. To ensure consistency and a fair comparison, all experiments were conducted with the temperature set to 0, and each experiment was repeated three times to compute an average, thereby eliminating any potential disruptions.

\begin{table*}[ht]
    \centering
    \small
    \renewcommand{\arraystretch}{1.1}
    \resizebox{\textwidth}{!}{%
    \begin{tabular}{ll cc cc cc}
        \toprule
        \multirow{2}{*}{\textbf{Models}} & \multirow{2}{*}{\textbf{Methods}} & \multicolumn{2}{c}{\textbf{EvalPerf}} & \multicolumn{2}{c}{\textbf{Mercury}} & \multicolumn{2}{c}{\textbf{ENAMEL}} \\
        \cmidrule(lr){3-4} \cmidrule(lr){5-6} \cmidrule(lr){7-8}
         &  & \textbf{DPS\_norm} & \textbf{Pass@1} & \textbf{Beyond@1} & \textbf{Pass@1} & \textbf{eff@1} & \textbf{Pass@1} \\
        \midrule
        \midrule

        \multirow{3}{*}{\shortstack{Qwen2.5-Coder\\-32B-Instruct}}
        & \tool       & \textbf{86.20} & \textbf{87.30} & \textbf{78.96} & 93.75 & \textbf{51.26} & \textbf{86.62} \\
         & w/o Uniqueness-1      & 80.84 \textcolor{darkred}{-5.36} & 79.66 \textcolor{darkred}{-7.64} & 76.75 \textcolor{darkred}{-2.21} & \textbf{94.14} \textcolor{darkgreen}{+0.39} & 50.95 \textcolor{darkred}{-0.31} & 80.98 \textcolor{darkred}{-5.64} \\
         & w/o Uniqueness-2      & 78.07 \textcolor{darkred}{-8.13} & 70.34 \textcolor{darkred}{-16.96} & 72.83 \textcolor{darkred}{-6.13} & 87.11 \textcolor{darkred}{-6.64} & 47.62 \textcolor{darkred}{-3.64} & 77.46 \textcolor{darkred}{-9.16} \\

        \midrule
\multirow{3}{*}{DeepSeek-V3} 
         & \tool       & \textbf{87.08} & \textbf{90.67} & \textbf{82.76} & \textbf{96.09} & \textbf{60.41} & \textbf{89.44} \\
         & w/o Uniqueness-1      & 80.91 \textcolor{darkred}{-6.17} & 85.59 \textcolor{darkred}{-5.08} & 81.79 \textcolor{darkred}{-0.97} & 95.31 \textcolor{darkred}{-0.78} & 54.70 \textcolor{darkred}{-5.71} & 85.92 \textcolor{darkred}{-3.52} \\
         & w/o Uniqueness-2      & 80.42 \textcolor{darkred}{-6.66} & 79.66 \textcolor{darkred}{-11.01} & 62.90 \textcolor{darkred}{-19.86} & 74.61 \textcolor{darkred}{-21.48} & 53.92 \textcolor{darkred}{-6.49} & 84.50 \textcolor{darkred}{-4.94} \\

        \bottomrule
    \end{tabular}
    }
    \caption{\textbf{\tool Uniqueness Study Results.} The results of \tool, \textbf{w/o Uniqueness-1}, and \textbf{w/o Uniqueness-2} are presented using Qwen2.5-Coder-32B-Instruct and DeepSeek-V3 as LLM backbones on the EvalPerf, Mercury, and ENAMEL benchmarks. We have highlighted the performance changes compared to \tool with colors.}
    \label{tab:special_results}
\end{table*}

\subsection{Main Results.}
We compare \tool with the other methods on the EvalPerf, Mercury, and ENAMEL benchmarks, and present the results in Table \ref{tab:complete_results}. First, we observe that direct instruction prompts yield good performance, indicating that LLMs have a reasonable understanding of correct and efficient code.  Then, through ECCO, we observe a slight improvement in efficiency on EvalPerf and Mercury. However, this improvement often comes at the cost of correctness. Particularly in more complex benchmarks like ENAMEL, the approach results in a decline in both efficiency and correctness. This suggests that relying solely on code understanding to generate optimization suggestions is insufficient. When optimization strategies are based purely on code-level analysis, they often fail to align with the broader logical requirements of the task. The mismatch between the code domain and the logic strategy domain makes such methods less effective.

Moreover, Effi-Learner shows some gains in efficiency and correctness on specific benchmarks, such as when using GPT-4o on the Mercury benchmark. However, its performance varies significantly across different LLM backbones and benchmarks, often falling short of the direct Instruct baseline. More importantly, Effi-Learner faces a recurring issue: both efficiency and correctness suffer simultaneously. This stems from its feedback mechanism, which focuses solely on performance metrics like execution time, neglecting the code’s functionality and correctness. Additionally, the lack of a comprehensive algorithmic strategy leads to an over-prioritization of execution time, often sacrificing code accuracy and resulting in a decline in both efficiency and correctness.

In comparison, \tool achieves a simultaneous improvement in both correctness and efficiency through efficiency optimizations at the logical and code implementation levels, followed by refinement of correctness using "checked" test case feedback. The results demonstrate that \tool delivers robust and consistent performance improvements across various benchmarks and different LLM backbones, with the gains highlighted in color. For example, using DeepSeek-V3 as the backbone on EvalPerf, \tool improved the efficiency metric DPS\_norm by 6.63\%, while on ENAMEL, eff@1 increased by 9.27\%, and Pass@1 improved by 2.82\%.

\subsection{Ablation Study.}
\tool incorporates several unique design choices, such as separating efficiency optimization into the logic domain and code implementation level. To better understand the impact of each component, we conduct the following ablation study:
\begin{itemize}[leftmargin=*,itemsep=2pt,topsep=0pt,parsep=0pt]
    \item \textbf{Variant-1:} (Without Algorithmic Exploration in the Logic Domain): In this variant, no algorithmic exploration is performed for efficiency optimization in the logic domain. Instead, the LLM directly generates the same count efficient code solution, followed by implementation-level optimization (based on the formalized task and the generated code solution). All other steps remain the same as in \tool.
    \item \textbf{Variant-2:} (Without Implementation Optimization in the Code Domain): This variant omits the implementation optimization step in the code domain, while all other processes are identical to those in \tool.
    \item \textbf{Variant-3:} (Without Code Correctness Refinement): In this variant, after generating the efficiency-optimized code solutions, the LLM independently selects the most efficient and correct code as the final output.
\end{itemize}
We conduct the ablation study using Qwen2.5-Coder-32B-Instruct and DeepSeek-V3 as LLM backbones, with the results presented in Table \ref{tab:ablation_results}. The results show that removing any component significantly impacts both efficiency and correctness. Specifically, omitting Algorithmic Exploration in the Logic Domain (\textbf{Variant-1}) or Implementation Optimization in the Code Domain (\textbf{Variant-2}) leads to a marked decline in efficiency metrics across all three benchmarks. Additionally, removing Code Correctness Refinement (\textbf{Variant-3}) results in a significant drop in Pass@1. These results align with our expectations, as both Algorithmic Exploration and Implementation Optimization are designed for efficiency, while Code Correctness Refinement ensures the final code retains functional correctness after efficiency-driven steps.

\section{Deeper Analysis}
\subsection{\tool Uniqueness Analysis.}
As mentioned in the Introduction, \tool has two distinct features: \textbf{\underline{Uniqueness 1:}} Separation of Efficiency Optimization into Logic and Code Domains, and \textbf{\underline{Uniqueness 2:}} The Order of Correctness and Efficiency. 
To gain a deeper understanding of these unique advantages, we conducted the following comparative experiments:
\begin{itemize}[leftmargin=*,itemsep=2pt,topsep=0pt,parsep=0pt]
    \item \textbf{w/o Uniqueness-1:} Rather than separating efficiency optimization into the logic and code domains, we prompt the LLM to generate code that is both efficient and correct. Then, based on the formalized task and the generated code, the LLM is queried to suggest any possible strategies for optimizing efficiency. Subsequently, we optimize the generated code according to these strategies, with the following steps remaining the same as in \tool. It is important to note that the difference between \textbf{w/o Uniqueness-1} and ECCO lies in the fact that ECCO provides correctness or efficiency strategies solely based on the code, whereas \textbf{w/o Uniqueness-1} generates efficiency-focused strategies based on both the formalized task and the generated code, followed by refinement for correctness.
    \item \textbf{w/o Uniqueness-2:} We changed the priority sequence of correctness and efficiency. In this approach, we first generate the code based on the formalized task and refine it for correctness. Then, we conduct algorithm exploration and implement optimal methods based on the formalized task and the refined code. Finally, we optimize the code using the explored algorithms and implementation suggestions.
\end{itemize}
The results of the uniqueness study are presented in Table \ref{tab:special_results}. As shown, in the absence of \textbf{Uniqueness-1}, the performance of Qwen2.5-Coder-32B-Instruct on Mercury showed a slight increase in Pass@1, but in other cases, the performance declined, particularly in terms of DPS\_norm on EvalPerf and eff@1 on ENAMEL. This underscores the importance of separating efficiency optimization into the logic and code domains. This separation effectively breaks down the challenge of optimizing code efficiency into manageable steps, making the overall optimization process more focused and targeted. On the other hand, in the absence of \textbf{Uniqueness-2}, both efficiency and correctness saw significant declines across all benchmarks and different LLM backbones. The main reason for this is that optimizing correctness before efficiency limits the LLM’s ability to explore efficient algorithms and practical optimizations. In fact, this approach often backfires, resulting in a situation where code correctness is sacrificed in the pursuit of efficiency. These findings further validate the "efficiency-first, correctness-later" strategy as a crucial approach for generating both efficient and correct code.

\begin{figure}
    \centering
    \includegraphics[width=0.50\textwidth]{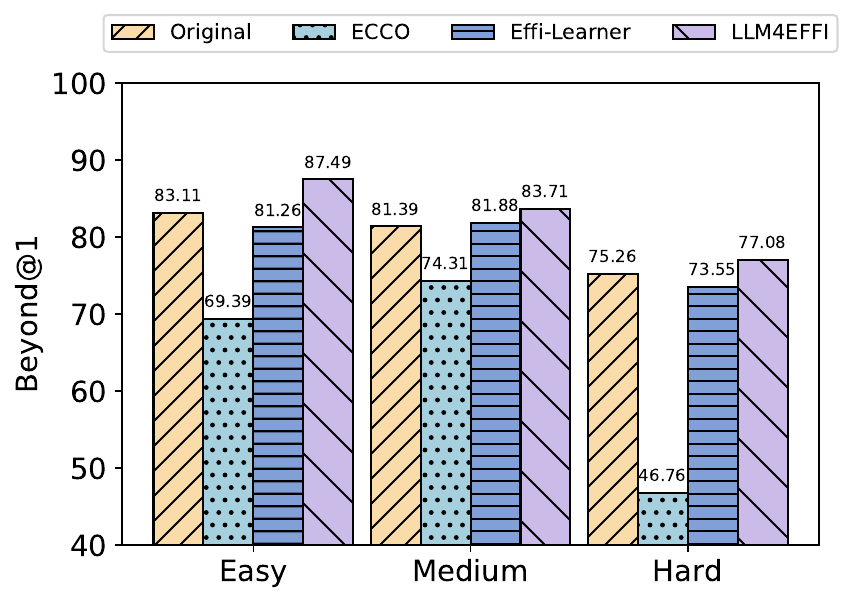}
    \caption{The Beyond@1 performance of \tool on tasks of varying difficulty levels in Mercury, with DeepSeek-V3 as the backbone.}
    \label{fig:Mercury}
\end{figure}

\subsection{Performance of Different Difficulty Levels.}
To evaluate the performance of \tool on tasks of varying difficulty levels, we conducted an analysis across three levels—Easy, Medium, and Hard—using Mercury with DeepSeek-V3 as the backbone. The results are shown in Figure \ref{fig:Mercury}. \tool consistently achieves the highest Beyond@1 metrics, outperforming other methods across tasks ranging from easy to difficult. This robust performance highlights the \tool’s effectiveness in tackling a broad spectrum of challenges. Additionally, we observed a significant drop in performance for ECCO on hard-level tasks. This decline is mainly due to the difficulty of providing valid optimization suggestions for complex code, a challenge that remains substantial for LLMs.

\subsection{Case Study.}
To provide a more intuitive demonstration of \tool, we conducted a case study, with the process detailed in Appendix \ref{appendix-casestudy}. It can be observed that methods like ECCO and Effi-Learner, which generate code first and then optimize for efficiency, are constrained by the algorithmic design and overall structure of the initial code, leading to only incremental improvements. In contrast, \tool breaks free from these constraints, enabling it to fully explore more efficient algorithms at a high level based on the task, while also incorporating practical level efficiency optimizations, thus achieving more effective efficiency optimization.

\section{Conclusion}
In this paper, we presented \tool, a novel framework designed to generate both efficient and correct code. By separating efficiency optimization into the logic and code domains and adopting an "efficiency-first, correctness-later" approach, \tool enables the exploration of a broader range of algorithmic solutions while maintaining functional correctness. Experimental results demonstrate \tool’s robust performance, with consistent improvements in both efficiency and correctness across different LLM backbones.

\section*{Limitation}
Although \tool excels at generating both efficient and correct code, it also has some limitations. One major challenge is the trade-off between efficiency and maintainability. In some cases, the generated efficient code may become more complex and harder to read. Achieving the right balance between efficiency and maintainability is not always straightforward, and in certain cases, highly efficient code may sacrifice readability and ease of future modifications. Future work will focus on optimizing the \tool to improve its scalability and extend its applicability to more complex software engineering tasks.

\bibliography{custom}
\appendix

\thispagestyle{empty}
\newpage

\definecolor{darkbrown}{RGB}{100, 80, 50}
\definecolor{darkorange}{RGB}{255, 140, 0}
\definecolor{darkred}{RGB}{139, 0, 0}
\definecolor{softblue}{RGB}{140, 160, 200}
\definecolor{lightgreen}{RGB}{145, 204, 117}
\definecolor{lightyellow}{RGB}{250, 200, 88}
\definecolor{lightred}{RGB}{238, 102, 102}
\definecolor{lightblue}{RGB}{115, 192, 222}

\newtcolorbox{promptbox}[2][Prompt]{
colback=black!5!white,
arc=5pt, 
boxrule=0.5pt,
fonttitle=\bfseries,
title=#1, 
before upper={\small}, fontupper=\fontfamily{ptm}\selectfont,
colframe=#2, 
}

\lstdefinestyle{pythonstyle}{
    language=Python,
    basicstyle=\ttfamily\footnotesize,
    keywordstyle=\color{blue},
    commentstyle=\color{green},
    stringstyle=\color{red},
    backgroundcolor=\color{lightgray!20},
    frame=single,
    breaklines=true,
    postbreak=\mbox{\textcolor{red}{$\hookrightarrow$}\space}
}

\renewcommand{\lstlistingname}{File}

\lstset{
    autogobble,
    columns=fullflexible,
    showspaces=false,
    showtabs=false,
    breaklines=true,
    showstringspaces=false,
    breakatwhitespace=true,
    breaklines=true,
    backgroundcolor=\color{lightgray!20},
    escapeinside={(*@}{@*)},
    commentstyle=\color{greencomments},
    keywordstyle=\color{bluekeywords},
    stringstyle=\color{redstrings},
    numberstyle=\color{graynumbers},
    basicstyle=\ttfamily\footnotesize,
    frame=l,
    framesep=12pt,
    xleftmargin=12pt,
    tabsize=4,
    captionpos=b
}

\section{Appendix of Prompts.}
\label{sec:ap-method}

\subsection{Prompts of \tool.}
\label{sec:eg-feature-extraction}

\begin{figure}[H]
\centering
\begin{promptbox}[Task Formalization]{softblue}
\textbf{System:} \\
As a professional algorithm engineer, please analyze the given algorithm problem according to the following categories. Do not provide any example implementation:
\begin{itemize}
    \item Entry Point Function Name\item Input/Output Conditions\item Edge Cases and Parameter Types (Int, String, etc.)\item Expected Behavior
\end{itemize}
\textbf{User:} \\
The algorithm problem description is as follows:\\ <natural language description>
\end{promptbox}
\caption{Task Formalization.}
\end{figure}

\begin{figure}[H]
\centering
\begin{promptbox}[Task Formalization Check]{softblue}
\textbf{System:} \\
As an excellent algorithm engineer, please analyze whether the explanation of the problem matches the original requirements. If they are consistent, output “Yes”. If they are not consistent, output “No” and provide the reason, as shown below:
\{"Yes":"NULL"\} \\
\{"No":"The reason is"\}

\textbf{User:} \\
<natural language description> \\
<task description>
\end{promptbox}
\caption{Checking the Task Formalization Result.}
\end{figure}

\begin{figure}[H]
\centering
\begin{promptbox}[Synthesize Test Case Inputs]{softblue}
\textbf{System:} \\
As a tester, your task is to create comprehensive test inputs for the function based on its definition and docstring. These inputs should focus on edge scenarios to ensure the code’s robustness and reliability. Please output all test cases in a single line, starting with input. \\
\textbf{User:} \\
EXAMPLES: \\
Function: 
\begin{lstlisting}[breaklines=true]
from typing import *
def find_the_median(arr: List[int]) -> float:
    Given an unsorted array of integers `arr`, find the median of the array.
    The median is the middle value in an ordered list of numbers.
    If the length of the array is even, then the median is the average of the two middle numbers.
\end{lstlisting}
Test Inputs (OUTPUT format): \\
input: [1] \\
input: [-1, -2, -3, 4, 5] \\
input: [4, 4, 4] \\
input: [....] \\
input: [....] \\
END OF EXAMPLES. \\
Function: \\
<task description>
\end{promptbox}
\caption{Synthesize Test Case Inputs.}
\end{figure}

\begin{figure}[H]
\centering
\begin{promptbox}[Implementation Optimization in Code Domain]{softblue}
\textbf{System:} \\
As a professional Python algorithm programming expert, please provide suggestions for improving code efficiency based on the potential inefficiencies mentioned above. For example: \\
1. Using xxx instead of xxx can significantly improve code efficiency. \\
Please provide at least 20 suggestions.\\
\textbf{User:} \\
<algorithm description>
\end{promptbox}
\caption{Implementation Optimization in Code Domain.}
\end{figure}

\begin{figure}[H]
\centering
\begin{promptbox}[Complete Test Case Generation]{softblue}
\textbf{System:} \\
As a programmer, your task is to calculate all test outputs and write the test case statement corresponding to the test input for the function, given its definition and docstring. Write one test case as a single-line assert statement. \\
\textbf{User:} \\
EXAMPLES: \\
Function: 
\begin{lstlisting}[breaklines=true]
from typing import List
def find_the_median(arr: List[int]) -> float:
    Given an unsorted array of integers `arr`, find the median of the array. The median is the middle value in an ordered list of numbers.
    If the length of the array is even, then the median is the average of the two middle numbers.
\end{lstlisting}
Test Input: \\
input: [1, 3, 2, 5]\\
Test Case: 
\begin{verbatim}
assert find_the_median([1, 3, 2, 5]) == 2.5
\end{verbatim}
END OF EXAMPLES. \\
FUNCTION: \\
<task description>
<input case>
\end{promptbox}
\caption{Complete Test Case Generation.}
\end{figure}

\begin{figure}[H]
\centering
\begin{promptbox}[Algorithmic Exploration in Logic Domain]{softblue}
\textbf{System:} \\
As a professional algorithm engineer, you can effectively design multiple algorithms to solve the problem with low time complexity and output them in pseudo algorithm format. A pseudo algorithm is a nonlinear, high-level programming language for algorithmic logic. It combines natural language and programming structures to express the steps and sums of algorithms. The main purpose of process algorithms is to clearly display the core ideas and logic of the algorithm without relying on specific programming language syntax. Please design 5 excellent algorithm solutions based on the problem description provided. The time complexity of the algorithm needs to be as small as possible, and try to output 5 algorithms in the form of a pseudo-algorithm in the following format:
PS: DO NOT provide implementation examples!
\begin{lstlisting}[breaklines=true]
```algorithm1
{algorithm key description: this algorithm using xxx, the key is to make sure xxx}
{pseudo algorithm: ..}

{algorithm key description: this algorithm using xxx, the key is to make sure xxx}
{pseudo algorithm: ..}

{algorithm key description: this algorithm using xxx, the key is to make sure xxx}
{pseudo algorithm: ..}

{algorithm key description: this algorithm using xxx, the key is to make sure xxx}
{pseudo algorithm: ..}

{algorithm key description: this algorithm using xxx, the key is to make sure xxx}
{pseudo algorithm: ..}
\end{lstlisting}
\textbf{User:} \\
<task description>
\end{promptbox}
\caption{Algorithmic Exploration in Logic Domain.}
\end{figure}

\begin{figure}[H]
\centering
\begin{promptbox}[Code Candidates Generation]{softblue}
\textbf{System:} \\
As a professional algorithm engineer, please convert the selected algorithm into corresponding code. Ensure the code is complete and well-formatted. When converting to a standardized format, be sure to follow the guidelines specified in the “original question format”: \\
1. Use the same function name as given in the original question format; do not rename it. \\
2. You may incorporate practical optimization details drawn from the knowledge base. \\
The final output format should be as follows:
\begin{verbatim}
```python
{<code>
```
\end{verbatim}
\textbf{User:} \\
<task description> \\
<algorithm description> \\
<efficiency optimization suggestions>
\end{promptbox}
\caption{Code Candidates Generation.}
\end{figure}

\begin{figure}[H]
\centering
\begin{promptbox}[Code Refinement for Correctness]{softblue}
\textbf{System:} \\
As a professional code programming algorithm expert, your task is to correct the code and ensure that the code is fixed without impacting its time complexity or practical efficiency. Then I will provide you with specific code and test cases. \\
Important Notes: \\
1. Do not alter the algorithm itself \\
2. Do not change the format, such as the function name. \\
3. Please output in the specified format. \\
4. Ensure there are no syntax errors. \\
Please output in this format:
\begin{verbatim}
```python
{code}
```
\end{verbatim}

\textbf{User:} \\
<task description> \\
<algorithm description> \\
<efficiency optimization suggestions>
\end{promptbox}
\caption{Code Refinement for Correctness.}
\end{figure}

\begin{figure}[H]
\centering
\begin{promptbox}[Final Results Selection on Code Candidates]{softblue}
\textbf{System:} \\
As a professional algorithm engineer, please help me choose the most efficient code from the following codes. It is worth mentioning that it is necessary to consider the time complexity and practical level comprehensively: \\
INPUT: \\
\{
"1":"def ...()....", \\
"2": "def ...()..." \\
\} \\
OUTPUT:
\begin{verbatim}
```text
{key}
```
\end{verbatim}
EXAMPLE: \\
INPUT: \\
\{
"1":"def ...()....", \\
"2": "def ...()..." \\
\} \\
OUTPUT:
\begin{verbatim}
```text
1
```
\end{verbatim}
\textbf{User:} \\
<corrected code candidate>
\end{promptbox}
\caption{Final Results Selection on Code Candidates (Optional).}
\end{figure}

\begin{figure}[H]
\centering
\begin{promptbox}[Direct Code Generation Prompt for Variant-1]{softblue}
\textbf{System:} \\
As a professional Python algorithm engineer, please solve the algorithms problem and generate a solution code. The final output format should be as follows:
\begin{verbatim}
```python
{code}
```
\end{verbatim}
\textbf{User:} \\
<task description>
\end{promptbox}
\caption{Direct Code Generation Prompt for Variant-1.}
\end{figure}

\begin{figure}[H]
\centering
\begin{promptbox}[Direct Code Generation Prompt for w/o Uniqueness-1\&w/o Uniqueness-2]{softblue}
\textbf{System:} \\
As a professional Python algorithm engineer, please solve the algorithm problem and generate 5 solution codes. Please improve the efficiency of the code as much as possible while ensuring the correctness of the code. The final output format should be as follows:
\begin{verbatim}
```python1
{code}
```
```python2
{code}
```
```python3
{code}
```
```python4
{code}
```
```python5
{code}
```
\end{verbatim}
\textbf{User:} \\
<task description>
\end{promptbox}
\caption{Direct Code Generation Prompt for w/o Uniqueness-1\&w/o Uniqueness-2.}
\end{figure}


\subsection{Prompts of Effi-Learner.}
\begin{figure}[H]
\centering
\begin{promptbox}[Original Code Generation Prompt in Effi-Learner]{softblue}
Please complete Python code based on the task description.\\
\# Task description:<Task description>\\
\#Solution:
\end{promptbox} 
\caption{Original Code Generation Prompt in Effi-Learner.}
\end{figure}

\begin{figure}[H]
\centering
\begin{promptbox}[Efficiency Optimization Prompt in Effi-Learner.]{softblue}
Optimize the efficiency of the following Python code based on the task, test case, and overhead analysis provided. Ensure the optimized code can pass the given test case.\\
Task Description:\\
<task description>\\
Test Case:\\
<test case>\\
Original Code:\\
\begin{verbatim}
```python
<original code>
```
\end{verbatim}
Overhead Analysis:\\
<profile of original code>\\
Optimization Rules:\\
- Encapsulate the optimized code within a Python code block (i.e., python[Your Code Here]).\\
- Do not include the test case within the code block.\\
- Focus solely on code optimization; test cases are already provided.\\
- Ensure the provided test case passes with your optimized solution.\\
\end{promptbox} 
\caption{Efficiency Optimization Prompt in Effi-Learner.}
\end{figure}

\subsection{Prompts of ECCO.}

\begin{figure}[H]
\centering
\begin{promptbox}[Original Code Generation Prompt in ECCO]{softblue}
Write a python code which is efficient in terms of runtime and memory usage for the following problem description. Wrap the optimized code in a block of 3 backticks
\end{promptbox} 
\caption{Original Code Generation Prompt in ECCO.}
\end{figure}

\begin{figure}[H]
\centering
\begin{promptbox}[Feedback Generation Prompt in ECCO]{softblue}
Give feedback in english for why the code solution below is incorrect or inefficient and how the program can be fixed based on the problem description.\\
<original code>
\end{promptbox} 
\caption{Feedback Generation Prompt in ECCO.}
\end{figure}

\begin{figure}[H]
\centering
\begin{promptbox}[Refine Prompt in ECCO]{softblue}
Refine the given incorrect or sub-optimal code solution based on the feedback specified below. Wrap the refined code in a block of 3 backticks\\
<optimization suggestion>\\
<original code>\\
\end{promptbox} 
\caption{Refine Prompt in ECCO.}
\end{figure}

\subsection{Prompts of Instruct.}

\begin{figure}[H]
\centering
\begin{promptbox}[Prompt for \textbf{Instruction} Baseline]{softblue}
Please generate an efficient and correct code directly
\end{promptbox} 
\caption{Prompt for \textbf{Instruction} Baseline.}
\end{figure}

\newpage
\section{Case Study.}
\label{appendix-casestudy}
\subsection{The Execution Details of Each Process of \tool.}
\begin{figure*}[t]
    \centering
    \includegraphics[width=1.0\textwidth]{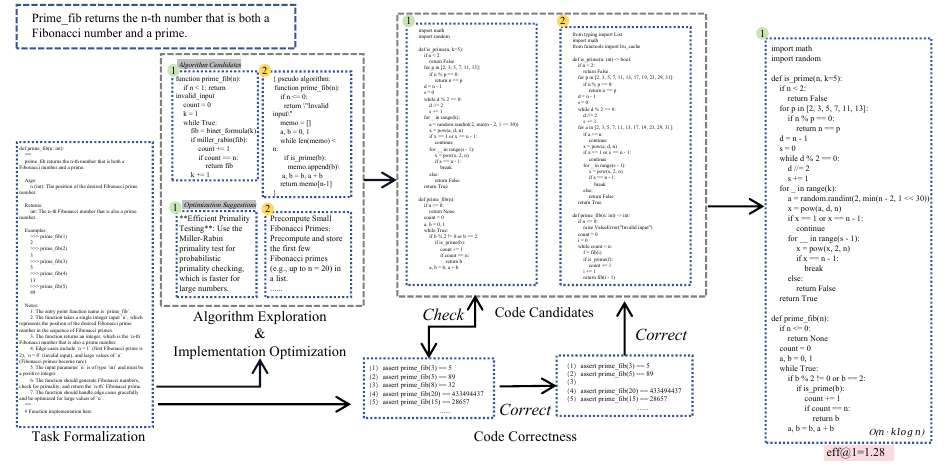}
    \caption{The figure illustrates the specific output of each subtask process of \tool in solving algorithm problems.}
    \label{fig:process_case}
\end{figure*}

As shown in Figure~\ref{fig:process_case}, \tool firstly analyzes the algorithm problem, "returns the n-th number that is both a Fibonacci number and a prime number", providing a detailed explanation of key aspects, including the entry point, expected behavior, and edge cases. Based on this analysis and the problem description, \tool explores potential algorithms and generates five efficient solutions, such as using the Fibonacci sequence generation method and Binet’s formula. Next, \tool examines the implementation details of these algorithms and identifies the optimal practical approaches. For example, it uses Python’s built-in pow() function for efficient exponentiation and applies the Miller-Rabin primality test (based on the Monte Carlo method) to enhance the efficiency of prime number detection for large numbers.

Then, \tool combines the explored algorithms and practical operations to generate five distinct code implementations. To validate the correctness of these codes, \tool generates 20 test cases based on the algorithm description and outputs them in the format "assert prime\_fib(3) == 5". Each code is then executed with these 20 test cases, recording the number of passed test cases ($Pass_{\text{t}} \leq 20$) and the number of successful executions for each test case ($Pass_{\text{c}} \leq 5$). Subsequently, \tool checks the test cases that are not passed by the code implementations, ensuring that correct test cases are not excluded due to code errors and preventing incorrect test cases from being misused in subsequent iterations.

After filtering, \tool obtains a new batch of test cases and executes them again to gather new results. For the failed test cases, an iterative feedback mechanism is applied to optimize the code. Then, the code, enhanced with the iterative feedback, is executed once more, and the final passing results are recorded. All codes are then ranked in descending order based on their correctness, and the most accurate code is selected. 

This process ensures the identification of the most optimal solution while maintaining both high efficiency and accuracy in code implementation.
\subsection{Comparison of Methods.}
\begin{figure*}[htbp]
    \centering

    \includegraphics[width=1.0\textwidth]{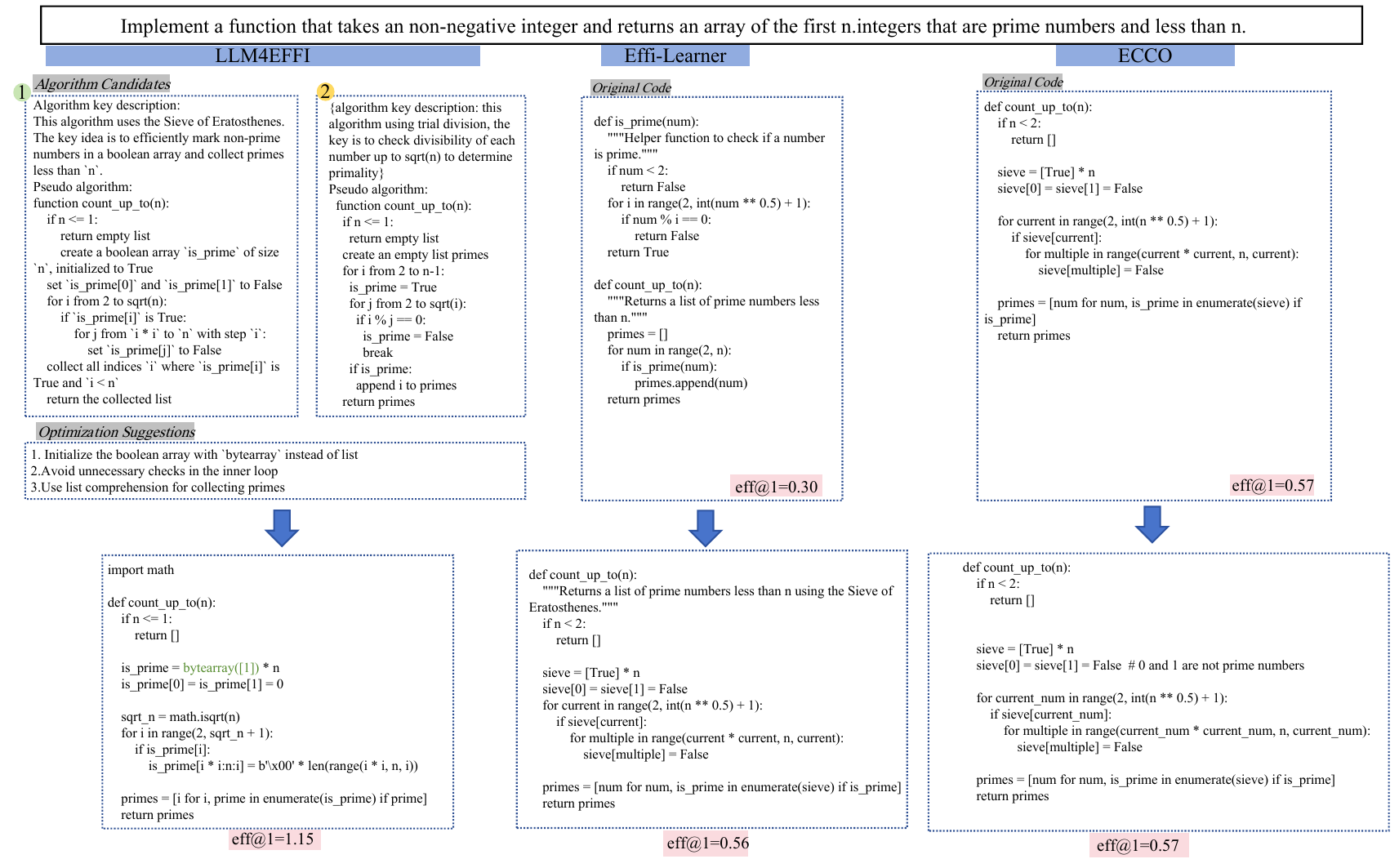}
    \caption{The diagram demonstrates how \tool, Effi-Learner, and ECCO generate code. \tool, through deep exploration of the algorithm domain, generates a set of efficient and high-quality algorithm candidates. However, the time complexity of these algorithms is similar, and there is no significant difference from the original code generated by Effi-Learner and ECCO. Subsequently, \tool identifies key optimization suggestions in its practical recommendations, such as replacing list with bytearray, among others. As a result, although the final code has a similar time complexity to the other two tools, it significantly outperforms them in the final ENAMEL efficiency evaluation metrics.}
    \label{fig:case2}
\end{figure*}

\begin{figure*}[htbp]
    \centering
    \includegraphics[width=1.0\textwidth]{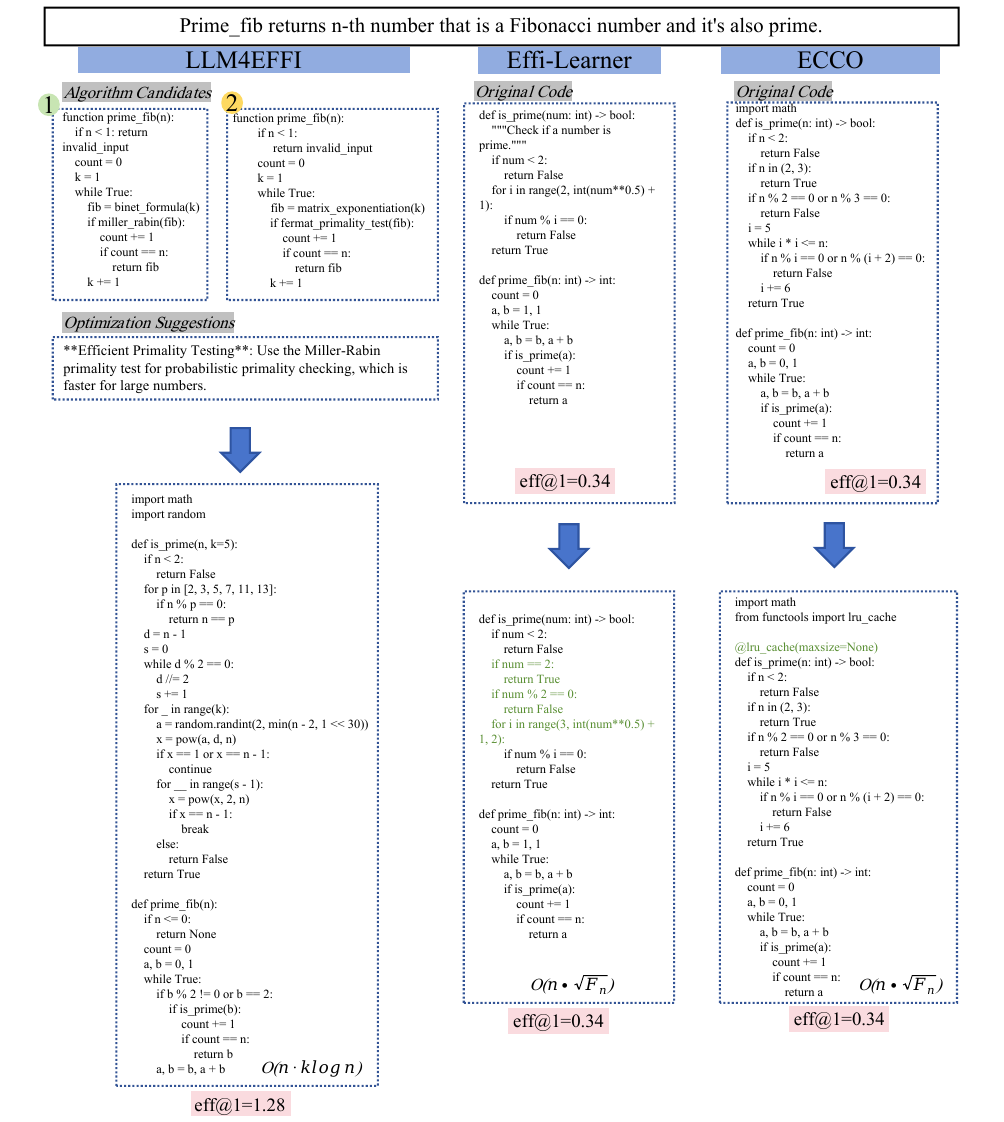}
    \caption{The figure illustrates the code generation process of \tool, Effi-Learner, and ECCO. \tool, through deep exploration of the algorithm domain, generates a set of efficient and high-quality algorithm candidates. By incorporating practical optimization suggestions, it ultimately produces an algorithm with a time complexity of only $\mathcal{O}(n \cdot k\log n)$, achieving a high score of 1.28 on the ENAMEL test set. In contrast, Effi-Learner and ECCO, constrained by the $\mathcal{O}(n \cdot \sqrt{F_n})$ time complexity of their code algorithms, can only perform local optimizations on certain implementations, resulting in minimal improvements, with the final efficiency index reaching only 0.34.}
    \label{fig:case1}
\end{figure*}

In Figures~\ref{fig:case2} and Figures~\ref{fig:case1}, we compare the code efficiency optimization processes of the three tools.

\end{document}